\documentclass[12pt]{iopart}
\usepackage{amsfonts}
\usepackage{graphicx}
\usepackage{epsfig}

\begin{document}

\title{Amplitude modulated Bloch oscillations of photon probability
distribution in cavity-atom system}
\author{G Zhang$^1$, W H Hu$^{1,2}$, Z Song$^1$}
\address{$^1$School of Physics, Nankai University, Tianjin 300071,
China}
\address{$^2$Beijing Computational Science Research Center,
Beijing 100084, China}
\ead{songtc@nankai.edu.cn}

\begin{abstract}
We study the dynamics of the Rabi Hamiltonian in the medium coupling regime
with $\left\vert g/\omega \right\vert \sim 0.07$, where $g$ is atom-field
coupling constant, $\omega $ is the field frequency, for the quantum state
with average photon number $\bar{n}\sim 10^{4}$. We map the original
Hamiltonian to an effective one, which describes a tight-binding chain
subjected to a staggered linear potential. It is shown that the photon
probability distribution of a Gaussian-type state exhibits the amplitude
modulated Bloch oscillations (BOs), which is a superposition of two
conventional BOs with a half-BO-period delay between them and is essentially
another type of Bloch-Zener oscillation. The probability transition between
the two BOs can be controlled and suppressed by the ratio $g\sqrt{\bar{n}}%
/\omega $, as well as in-phase resonant oscillating atomic frequency $\Omega
\left( t\right) $, leading to multiple zero-transition points.
\end{abstract}

\maketitle

%{03.67.Hk, Quantum communication}
%{03.67.Lx, Quantum computation}
%{03.65.Xp Tunneling, traversal time, quantum Zeno dynamics }
%{42.50.-p, Quantum optics}
%{42.50.Dv, Nonclassical states of the electromagnetic field, including entangled photon states; quantum state engineering and measurements}
%{03.65.wj, State reconstruction, quantum tomography}
%{03.67.-a, Quantum information}
%{42.50.ex, Optical implementations}
%{32.80.Pj, Optical cooling of atoms; trapping}
%{42.50.Hz, Strong-?eld excitation of optical transitions in quantum systems; multi-photon processes; dynamic Stark shift}
%{42.50.Pq, Cavity quantum electrodynamics; micromasers }

\section{Introduction}

\label{sec_intro}

The interaction of matter and light is one of the fundamental processes
occurring in nature. Its elemental constituent is the coupling between
single atoms and photons. The quantum Rabi model considers a two-level atom
coupled to a quantized field, describing the simplest interaction between
quantum light and matter. It was used to describe the interaction between a
rapidly varying, weak magnetic field and nuclear spin \cite%
{I.I.Rabi36,I.I.Rabi37}. Now it applies to a variety of physical systems
with the two-level atom, being trapped ions, quantum dots, and
superconducting qubits. Such systems can be exploited as a building block
for quantum information processing and other potential applications to
future quantum technologies \cite{M.A.Nielsen04}.

Although the quantum Rabi model seems relatively simple and was declared to
be solved recently \cite{D.Break11,E.Solano11}, the dynamics are actually
quite complicated, depending on the system parameters and the initial state
\cite{J.Casanova}. The rotating wave approximation (RWA) is justified in the
strong coupling regime (but the average photon number can not be too large)
and results in the Jaynes-Cummings (JC) model \cite{E.T.Jaynes63}, which
predicts non-classical phenomena, such as revivals of the initial excited
state of the atom \cite{P.Meystre75,J.H.Eberly80,N.B.Narozhny}. It is well
known that an atom-cavity system undergoes the Rabi oscillations. In the
case of the single cavity mode being initially prepared in a coherent state
with large average photon number, although the oscillations experience the
collapse-and-revival, the photon probability distribution remains unchanged
in the framework of the JC model. The question is whether it is true if the
original Rabi model is considered, or what will be observed in experiment.
With strong couplings in the solid-state-cavity-QED \cite{Nature.450.857},
the failure of RWA has relighted the interest on the quantum Rabi model%
\textbf{\ }\cite{D.Break11, J.Casanova, AnnPhys.16.767, PhysRevB.72.115303,
JPhysBAtMolOptPhys.43.175501, PhysRevA.86.023822,
JPhysA.MathTheor.46.335301, Ridolfo12, Felicetti14, G14}.

In this paper, we study the dynamics of the Rabi Hamiltonian in the medium
coupling regime with $\left\vert g/\omega \right\vert \sim 0.07$, where $g$
is atom-field coupling constant, $\omega $\ is the field frequency, for the
quantum state with average photon number $\bar{n}\sim 10^{4}$. The effective
coupling constant becomes $\left\vert g\sqrt{\bar{n}}\right\vert \sim
7\omega $, which goes beyond the RWA. Within this parameter regime, the
original Hamiltonian is mapped to an effective one, which describes a
uniform tight-binding chain subjected to a staggered linear potential. An
approximate solution suggests that the photon number distribution of a
Gaussian-type state exhibits the amplitude modulated Bloch oscillations
(BOs), which is a superposition of two conventional BOs with a
half-BO-period delay between them. This phenomenon essentially belongs to
Bloch-Zener oscillation in the strong tunneling limit \cite{B.M.Breid,
NewJPhys.9.62, PhysRevA.83.013609, EuroPhysLett.76.416,
PhysRevLett.101.193902, PhysRevB.85.115433}. We find that the probability
transition between the two BOs can be controlled and suppressed by the ratio
$g\sqrt{\bar{n}}/\omega $, as well as resonant oscillating atomic frequency $%
\Omega \left( t\right) $, leading to multiple zero-transition\ points. A
numerical simulation of dynamics in the Rabi model confirms this prediction.

This paper is organized as follows. In Section \ref{sec_model}, we propose
the effective Hamiltonian in the concerned parameter regime. In Section \ref%
{sec_approximate formalism}, we present the approximate solution of the
effective Hamiltonian, based on which the dynamics of a local state is
investigated. Section \ref{sec_Controllable dynamics} is dedicated to a
numerical simulation in the original Rabi model, focusing the control of the
probability transition between two BOs by the atomic transition frequency.
Finally, we give a summary and discussion in Section \ref{sec_Summary}.

\section{Model and equivalent Hamiltonian}

\label{sec_model}We start by considering the single-mode atom-cavity model
whose Hamiltonian can be written as
\begin{equation}
H=\omega \hat{a}^{\dag }\hat{a}+\frac{\Omega }{2}\hat{\sigma}_{x}+g\hat{%
\sigma}_{z}(\hat{a}^{\dag }+\hat{a}),  \label{H_tot}
\end{equation}%
where $\omega $ and $\Omega $ are the field and atomic transition
frequencies, respectively, and $g$ is the coupling constant. $\hat{a}^{\dag
} $ ($\hat{a}$) is the creation (annihilation) operator of the light field,
while $\hat{\sigma}_{x}=\left\vert e\right\rangle \left\langle e\right\vert
-\left\vert g\right\rangle \left\langle g\right\vert $, $\hat{\sigma}%
_{z}=\left\vert e\right\rangle \left\langle g\right\vert +\left\vert
g\right\rangle \left\langle e\right\vert $ are atomic operators, where $%
\left\vert g\right\rangle $ and $\left\vert e\right\rangle $ denote the
ground and excited atomic states, respectively. There is an approximation
which has been developed, the rotating-wave approximation (RWA) \cite%
{Jaynes63,Shore93} under the condition $\left\vert g\sqrt{\bar{n}}%
\right\vert \ll \omega $.

In this paper, we consider the case with the medium coupling regime with $%
\left\vert g/\omega \right\vert \sim 0.07$ and the average photon number $%
\bar{n}\sim 10^{4}$. Such a parameter regime is accessible in experiments
\cite{T.Niemczyk,P.Forn,K.D.B.Higgins}. The aim of this paper is to
investigate the dynamics of a particular initial state with a Gaussian-type
photon number distribution, which allows us to obtain the approximate
analytical result from the Hamiltonian (\ref{H_tot}).

\begin{figure}[tbp]
\begin{center}
\includegraphics[bb=60 250 520 760, width=0.45\textwidth, clip]{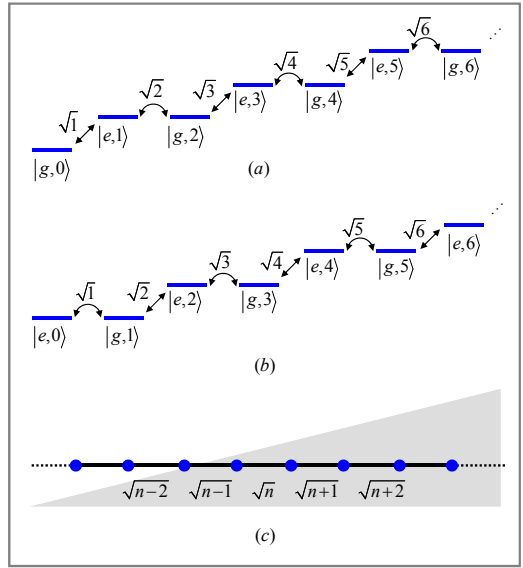}
\end{center}
\caption{(Color online) The atom-cavity level structure used for the mapping
to a tight-binding chain, and the implementation of Bloch oscillations in
the photon-number space. Atom-cavity level diagram showing the lower-energy
states for a two-level atom of transition frequency $\Omega $ coupled (with
single-photon Rabi frequency $g$) to a single mode cavity field of frequency
$\protect\omega $, with $\Omega =\protect\omega $. (a) and (b)\ correspond
to even- and odd-parity cases, respectively.\ The transition strengths are
in units of $g$. (c) For large photon region $n\gg 1$, both even- and
odd-parity cases can be mapped to a tight-binding chain with uniform NN
hopping strength ($\protect\sqrt{n+1}\simeq \protect\sqrt{n}$) and linear
potential (as represented by gray shadow)\ approximately, which allows the
occurrence of Bloch oscillations.}
\label{figure1}
\end{figure}

Since Hamiltionian (\ref{H_tot}) is parity-conserving for the excitation
number%
\begin{equation}
\mathcal{N}=a^{\dag }a+\frac{1}{2}\left( \hat{\sigma}_{x}+1\right) ,
\label{excitation number}
\end{equation}%
it can be written in two independent equivalent Hamiltonian $H_{\mathrm{eq}%
}^{\mathrm{e}}$ ($H_{\mathrm{eq}}^{\mathrm{o}}$) in the bases with even
(odd) excitation number:%
\begin{eqnarray}
H_{\mathrm{eq}}^{\lambda } &=&g\sum_{i=0}^{\infty }\sqrt{i+1}\left(
\left\vert i\right\rangle _{\lambda }\left\langle i+1\right\vert +\mathrm{%
H.c.}\right)  \nonumber \\
&&+\sum_{i=0}^{\infty }\left[ \frac{\left( -1\right) ^{\gamma _{\lambda }+i}%
}{2}\Omega +i\omega \right] \left\vert i\right\rangle _{\lambda
}\left\langle i\right\vert ,  \label{H_eq}
\end{eqnarray}%
where $\lambda =\mathrm{e,\ o}$ and $\gamma _{\mathrm{e}}=1$, $\gamma _{%
\mathrm{o}}=0$. This is schematically illustrated in Fig. \ref{figure1}. The
equivalence between $H$ and $H_{\mathrm{eq}}^{\lambda }$ is based on the
mapping of the corresponding basis: for even invariant subspace we have $%
\left\{ \left\vert \mathrm{g},0\right\rangle ,\left\vert \mathrm{e}%
,1\right\rangle ,\left\vert \mathrm{g},2\right\rangle ,\left\vert \mathrm{e}%
,3\right\rangle ,\left\vert \mathrm{g},4\right\rangle ,\cdots \right\} $ $%
\longrightarrow \left\{ \left\vert i\right\rangle _{\mathrm{e}}\right\} $,
where $i=0,1,2,\cdots $, while for odd invariant subspace we have $\left\{
\left\vert \mathrm{e},0\right\rangle ,\left\vert \mathrm{g},1\right\rangle
,\left\vert \mathrm{e},2\right\rangle ,\left\vert \mathrm{g},3\right\rangle
,\left\vert \mathrm{e},4\right\rangle ,\cdots \right\} $ $\longrightarrow
\left\{ \left\vert i\right\rangle _{\mathrm{o}}\right\} $ \cite{J.Casanova}.
Note that $H_{\mathrm{eq}}^{\lambda }$ is a standard tight-binding chain
with coordinate-dependent nearest neighbor (NN) hopping strength and on-site
potential. It is easy to find out that, within the region of large
excitation-number limit $i\gg 1$, when the initial state is local near $\bar{%
n}\sim 10^{4}$, we can take truncated approximation to $H_{\mathrm{eq}%
}^{\lambda }$. In this case, the expansion of the coupling strength in
equation (\ref{H_eq}) is $\sqrt{i}=\sqrt{\bar{n}}+\frac{i-\bar{n}}{2\sqrt{%
\bar{n}}}-\frac{\left( i-\bar{n}\right) ^{2}}{8\bar{n}^{3/2}}+...$. We take
the first term only, then $\sqrt{i}\approx \sqrt{\bar{n}}$; therefore, $H_{%
\mathrm{eq}}^{\lambda }$ can be approximately equivalent to a tight-binding
chain with uniform NN hopping strength and staggered linear potential. In
the following, we neglect the label $\lambda $\ in the effective Hamiltonian
for simplicity and demonstrate that such a discrete system admits the
existence of a robust BO \cite{H.Fukuyama, T.Hartmann, PhysRep.366.103}.

\section{Approximate formalism}

\label{sec_approximate formalism}In this section, we will introduce an
approximate formalism for the approximate solution of the effective
Hamiltonian and investigate the dynamical behavior as an application.

\subsection{Effective Hamiltonian}

We rewrite the effective Hamiltonian in the form%
\begin{eqnarray}
H_{\mathrm{eff}} &=&H_{\mathrm{0}}+H_{\mathrm{1}}  \label{H_eff_1} \\
H_{\mathrm{0}} &=&g\sqrt{\bar{n}}\sum_{i=0}^{N}\left( \left\vert
i\right\rangle \left\langle i+1\right\vert +\mathrm{H.c.}\right) +\omega
\sum_{i=0}^{N}\left( i-\frac{N}{2}\right) \left\vert i\right\rangle
\left\langle i\right\vert , \\
H_{\mathrm{1}} &=&\frac{\Omega }{2}\sum_{i=0}^{N}\left( -1\right)
^{i}\left\vert i\right\rangle \left\langle i\right\vert ,
\end{eqnarray}%
where $\bar{n}$\ is the average number of photon and can be regarded as a
constant in the context of the problem we concern in this paper. This
Hamiltonian is equivalent to (\ref{H_eq}) for the dynamics of a local state
within the region $i\gg 1$. Notice that the hamiltonian $H_{\mathrm{eff}}$\
is nothing but the tight-binding Hamiltonian to describe a single particle
subjected to a staggered linear potential, which has been well studied in
previous literatures \cite{B.M.Breid}. It is demonstrated that the
Bloch-Zener oscillations with a single period occurs in the parameter regime
$\left\vert g/\omega \right\vert \sim 0.2$ and $\Omega =6.734\omega $. In
this paper, we revisit the same model within the parameter regime $%
\left\vert g/\omega \right\vert \sim 0.07$\ and $\Omega =\omega $, which is
accessible in experiment \cite{T.Niemczyk,P.Forn,K.D.B.Higgins}. We will
show that the dynamics undergoes Bloch-Zener oscillations\ characterized by
two periods.

In the absence of term $H_{\mathrm{1}}$, it is a standard model which admits
the existence of Wannier-Stark localization and Bloch oscillations, and has
been studied extensively \cite{F.Bloch,C.Zener,V.G.Lyssenko,M.Cristiani}.
Now the question is: what happens to the dynamics of the model in Eq. (\ref%
{H_eff_1}). The basic idea is to consider the term $H_{\mathrm{1}}$\ as a
perturbation. According to the theory of Wannier-Stark localization \cite%
{H.Fukuyama, T.Hartmann}, the solution of $H_{\mathrm{0}}$\ has the form%
\begin{equation}
H_{\mathrm{0}}\left\vert \psi _{m}\right\rangle =E_{m}^{0}\left\vert \psi
_{m}\right\rangle ,
\end{equation}%
with%
\begin{equation}
E_{m}^{0}=\omega m-\frac{N}{2}\omega ,  \label{E_m}
\end{equation}%
and

\begin{eqnarray}
\left\vert \psi _{m}\right\rangle &=&\frac{1}{\sqrt{N}}\sum_{k}\exp \left[
-i\left( mk+\frac{L}{2}\sin k\right) \right] \left\vert k\right\rangle
\label{Psi0_alpha} \\
\left\vert k\right\rangle &=&\frac{1}{\sqrt{N}}\sum_{l=1}^{N}e^{\mathrm{i}%
kl}\left\vert l\right\rangle ,m=0,\pm 1,\pm 2,\ldots .
\end{eqnarray}%
In real space, it reads%
\begin{equation}
\left\vert \psi _{m}\right\rangle =\sum\limits_{l=1}^{2N}\mathrm{J}%
_{l-m}\left( \frac{L}{2}\right) \left\vert l\right\rangle .
\end{equation}%
where $L=-4g\sqrt{\bar{n}}/\omega $, is the spatial extent of a single Bloch
wave packet oscillation \cite{T.Hartmann}, and $g$ is negative.

In order to examine the effect of $H_{\mathrm{1}}$\ on the dynamics of the
system, it is convenient to work in the interaction picture. The propagator
of the whole Hamiltonian $H$ in the interaction picture is obtained by the
unitary transformation

\begin{equation}
H_{\mathrm{I}}\left( t\right) =e^{\mathrm{i}H_{0}t}H_{\mathrm{1}}e^{-\mathrm{%
i}H_{0}t}.
\end{equation}%
The propagator is%
\begin{equation}
U^{I}\left( t,0\right) =\mathcal{T}\exp \left[ -\mathrm{i}%
\int_{0}^{t}H_{I}\left( t^{\prime }\right) dt^{\prime }\right] ,  \label{U}
\end{equation}%
where $\mathcal{T}$\ is the time order operator. In the basis of states $%
\left\{ \left\vert \psi _{n}\right\rangle \right\} $, one obtains%
\begin{equation}
\left\langle \psi _{m}\right\vert H_{\mathrm{I}}\left( t\right) \left\vert
\psi _{n}\right\rangle =e^{\mathrm{i}\left( m-n\right) \omega t}\frac{\Omega
}{2}\left( -1\right) ^{m}\mathrm{J}_{n-m}\left( L\right) .
\end{equation}%
We will reduce it by the following two steps. Firstly, we can neglect
rapidly-oscillating terms in $e^{\mathrm{i}n\omega t}$\ for $n\geq 2$.
Secondly, we note that $\left\vert \mathrm{J}_{1}\left( x\right) \right\vert
\ll \left\vert \mathrm{J}_{0}\left( x\right) \right\vert $\ when $x$\ is
around the zeros of $\left\vert \mathrm{J}_{1}\left( x\right) \right\vert $,
$\left\vert \mathrm{J}_{1}\left( x_{0}\right) \right\vert =0$. Then when the
parameter $L\simeq x_{0},$ with $x_{0}=2.41$, $5.53$, $...$, one can drop
the term of $\left\langle \psi _{m}\right\vert H_{\mathrm{I}}\left( t\right)
\left\vert \psi _{m\pm 1}\right\rangle $.\ Therefore the approximation and
the feature of the Bessel function lead to the following approximate
expression%
\begin{equation}
\left\langle \psi _{m}\right\vert H_{\mathrm{I}}\left( t\right) \left\vert
\psi _{n}\right\rangle \approx \frac{\Omega }{2}\left( -1\right) ^{m}\mathrm{%
J}_{0}\left( L\right) \delta _{mn},
\end{equation}%
i.e., state $\left\vert \psi _{n}\right\rangle $\ is the simultaneous
eigenstate of time-dependent Hamiltonian $H_{\mathrm{I}}\left( t\right) $.
Then in the basis of states $\left\{ \left\vert \psi _{n}\right\rangle
\right\} $,\ the propagator in Eq. (\ref{U}) is reduced to
\begin{equation}
U_{mn}^{I}\left( t,0\right) \approx \delta _{mn}\exp \left[ -\mathrm{i}%
\int_{0}^{t}\left\langle \psi _{m}\right\vert H_{\mathrm{I}}\left( t^{\prime
}\right) \left\vert \psi _{n}\right\rangle dt^{\prime }\right] .
\label{U_Imn}
\end{equation}%
The corresponding propagator $U\left( t,0\right) $\ in Schr\"{o}dinger
picture is in the form
\begin{equation}
U_{mn}\left( t,0\right) \approx \delta _{mn}\exp \left\{ -\mathrm{i}\left[
m\omega +\left( -1\right) ^{m}\gamma \right] t\right\} ,  \label{U_mn}
\end{equation}%
where
\begin{equation}
\gamma =\frac{\Omega }{2}\mathrm{J}_{0}\left( L\right) .  \label{gamma}
\end{equation}

\subsection{Dynamics}

To clarify the feature of the dynamics, we consider the time evolution of an
arbitrary state. We note that any state can be decomposed into even and
odd-parity portions, which are spanned by the state $\left\vert \psi
_{n}\right\rangle $\ with even or odd $n$, respectively. It shows that the
parity is conserved during the time evolution under an arbitrary (either odd
or even) $H$ within the approximate framework. We see that the even and
odd-parity parts of quantum state evolve independently. Furthermore, the
appearance of the term $H$\ only contributes an overall phase on each parts.
It is presumable that the profile of the evolved state is determined by the
factor $\gamma $ in a simply manner. It plays a central role in the emerging
interference phenomena of two subwaves with different parity.

Actually, for a given initial state
\begin{equation}
\left\vert \psi \left( 0\right) \right\rangle =\sum_{n}f_{n}\left\vert \psi
_{n}\right\rangle ,
\end{equation}%
we have
\begin{eqnarray}
\left\vert \psi \left( t\right) \right\rangle &=&U\left( t,0\right)
\left\vert \psi \left( 0\right) \right\rangle  \nonumber \\
&=&e^{-\mathrm{i}\gamma t}\sum_{n}e^{-\mathrm{i}2n\omega t}f_{2n}\left\vert
\psi _{2n}\right\rangle +e^{\mathrm{i}\gamma t}\sum_{n}e^{-\mathrm{i}\left(
2n+1\right) \omega t}f_{2n+1}\left\vert \psi _{2n+1}\right\rangle .
\label{Psi(t)}
\end{eqnarray}%
In order to demonstrate the physical picture of state $\left\vert \psi
\left( t\right) \right\rangle $, we consider the time evolution of the same
initial state under the free Hamiltonian $H_{\mathrm{0}}$, which can be
written as%
\begin{equation}
\left\vert \psi _{0}\left( t\right) \right\rangle =e^{-\mathrm{i}%
H_{0}t}\left\vert \psi \left( 0\right) \right\rangle =\sum_{n}e^{-\mathrm{i}%
n\omega t}f_{n}\left\vert \psi _{n}\right\rangle .
\end{equation}%
It is turned out that $\left\vert \psi _{0}\left( t\right) \right\rangle $
exhibits standard Bloch oscillation with period $T_{B}=2\pi /\omega $\ for a
local initial state. Furthermore, a straightforward derivation shows that%
\begin{equation}
\left\vert \psi \left( t\right) \right\rangle =\cos \left( \gamma t\right)
\left\vert \psi _{0}\left( t\right) \right\rangle -\mathrm{i}\sin \left(
\gamma t\right) \left\vert \psi _{0}\left( t+\frac{T_{B}}{2}\right)
\right\rangle .
\end{equation}%
It shows that state $\left\vert \psi \left( t\right) \right\rangle $\ can be
regarded as the superposition of two evolved states under the Hamiltonian $%
H_{\mathrm{0}}$. In order to get a physical picture of the phenomenon, we
introduce an operator $\hat{B}_{\pi }$\ which relates two states $\left\vert
\psi _{0}\left( t\right) \right\rangle \ $and $\left\vert \psi _{0}\left( t+%
\frac{T_{B}}{2}\right) \right\rangle $\ in the following way%
\begin{equation}
\hat{B}_{\pi }\left\vert \psi _{0}\left( t\right) \right\rangle =\left\vert
\psi _{0}\left( t+\frac{T_{B}}{2}\right) \right\rangle .
\end{equation}%
Obviously, $\hat{B}_{\pi }$\ can be expressed in term of $H_{\mathrm{0}}$\
as
\begin{equation}
\hat{B}_{\pi }=e^{-\mathrm{i}H_{0}T_{B}/2}.
\end{equation}%
The physical meaning of operator $\hat{B}_{\pi }$\ becomes clear\ if we
apply it on the state $\left\vert k\right\rangle $. A direct derivation
shows that%
\begin{equation}
\hat{B}_{\pi }\left\vert k\right\rangle =\exp \left[ \mathrm{i}L\sin k\right]
\left\vert k+\pi \right\rangle .
\end{equation}%
It indicates that operator $\hat{B}_{\pi }$\ is nothing but the $\pi $-boost
operator that shifts all momentum states by $\pi $ with an extra phase
factor. Then if we consider the initial state\ as a Gaussian wave packet $%
\left\vert \phi \left( n_{0},k_{0}\right) \right\rangle $ with momentum $%
k_{0}$ and the center position $n_{0}$, approximately, it evolves as%
\begin{eqnarray}
\left\vert \psi \left( t\right) \right\rangle &=&\cos \left( \gamma t\right)
e^{-\mathrm{i}H_{\mathrm{0}}t}\left\vert \phi \left( n_{0},k_{0}\right)
\right\rangle  \nonumber \\
&&-\mathrm{i}\sin \left( \gamma t\right) e^{-\mathrm{i}H_{\mathrm{0}}t}\exp
\left( \mathrm{i}L\sin k_{0}\right) \left\vert \phi (n_{0}+n_{\pi
},k_{0}+\pi )\right\rangle  \label{key eq}
\end{eqnarray}%
where $n_{\pi }=-L\cos k_{0}$. Here the Gaussian wave packet has the form%
\begin{equation}
\left\vert \phi \left( n_{0},k_{0}\right) \right\rangle =\frac{1}{\sqrt{R}}%
\sum_{n=0}^{\infty }e^{-\frac{\alpha ^{2}}{2}\left( n-n_{0}\right) ^{2}}e^{%
\mathrm{i}k_{0}n}\left\vert n\right\rangle ,  \label{GWP}
\end{equation}%
where $R$ is the normalization factor and $\alpha $\ determines the half
width of the wave packet, $\Delta =2\sqrt{2\ln 2}/\alpha $ in the case of $%
\alpha \ll 1$. We can see that the evolved state represents the
superposition of Bloch oscillations of two wave packets with a $n_{\pi }$%
-shifted in position and a $\pi $-shifted in momentum. The amplitude of each
oscillation is modulated with sinusoidal time dependence.

\begin{figure}[tbp]
\begin{center}
\includegraphics[bb=0 200 550 620, width=0.45\textwidth, clip]{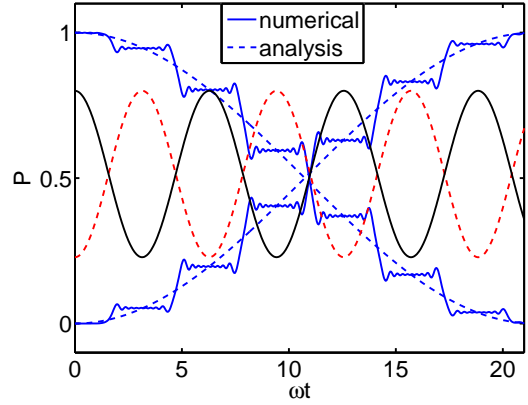}
\end{center}
\caption{(Color online) Probability transition between two BOs obtained by
approximate method presented in this paper and numerical simulation. The
initial wavepacket $\left\vert \protect\psi \left( 0\right) \right\rangle
=\left\vert \protect\phi \left( n_{0},0\right) \right\rangle $ with $%
n_{0}=1.01\times 10^{4}$ and $\protect\alpha =0.1$ is involved in the
Hamiltonian $H_{\mathrm{eff}}$\ with $\Omega =\protect\omega $ and $L =28.89$%
. The black solid line and red dashed line are the shapes of Eq. (\protect
\ref{n_ab}), while blue dashed lines represent the Eq. (\protect\ref{p_ab}).
The solid blue lines indicate the numerical results for $P_{a}\left(
t\right) $ and $P_{b}\left( t\right) $, which is in accord with our
prediction that $\Omega $ can lead to the probability transition in the
special intervals.}
\label{figure2}
\end{figure}

Particularly, we see that in the case of the parameters satisfying the
equation
\begin{equation}
\gamma nT_{B}=\frac{\pi }{2},\left( n\in
%TCIMACRO{\U{2115} }%
%BeginExpansion
\mathbb{N}
%EndExpansion
\right)
\end{equation}%
the wave packet $\left\vert \phi \left( n_{0},k_{0}\right) \right\rangle $\
is separated as two equal-probability ones, $\left\vert \phi \left(
n_{0},k_{0}\right) \right\rangle $ and its counterpart $\left\vert \phi
(n_{0}+n_{\pi },k_{0}+\pi )\right\rangle $, at instant $t=\frac{nT_{B}}{2}$.
On the other hand, we see that $\left\vert \psi \left( t\right)
\right\rangle $\ and $\left\vert \psi _{0}\left( t\right) \right\rangle $\
coalesce with each other at instants $t=t_{n}$, where%
\begin{equation}
t_{n}=\frac{n\pi }{\gamma },\left( n\in
%TCIMACRO{\U{2115} }%
%BeginExpansion
\mathbb{N}
%EndExpansion
\right) .
\end{equation}

To exemplify these features, we give the center positions $n_{a}\left(
t\right) $, $n_{b}\left( t\right) $ and corresponding probabilities $%
P_{a}\left( t\right) $, $P_{b}\left( t\right) $ of two wave packets as
function of time for the initial state $\left\vert \psi \left( 0\right)
\right\rangle =\left\vert \phi \left( n_{0},0\right) \right\rangle $ as the
form%
\begin{equation}
n_{a}\left( t\right) =n_{b}+L\cos \left( \omega t\right) =n_{0}-L\sin
^{2}\left( \omega t/2\right)  \label{n_ab}
\end{equation}%
and%
\begin{equation}
P_{a}\left( t\right) =1-P_{b}\left( t\right) =\cos ^{2}\left( \gamma t\right)
\label{p_ab}
\end{equation}%
which is evolved in the Hamiltonian $H_{\mathrm{eff}}$ in Eq. (\ref{H_eff_1}%
) with constant $\Omega $.

However, it is worthy to point out that the above conclusions are obtained
within the approximate framework, which we will see from the following
analysis. One can consider the mechanism of the modulated BOs in an
alternative way. For a wave packet $\left\vert \phi \left(
n_{0},k_{0}\right) \right\rangle $ at a certain location, the action of $H_{%
\mathrm{0}}$\ is to drive it to evolve as a Bloch oscillation, while the
action of $H_{\mathrm{1}}$\ is to make a transition from $\left\vert \phi
\left( n_{0},k_{0}\right) \right\rangle $\ to its counterpart, i.e.,

\begin{equation}
H_{\mathrm{1}}\left\vert \phi \left( n_{0},k_{0}\right) \right\rangle
\rightarrow \left\vert \phi (n_{0},k_{0}+\pi )\right\rangle .
\end{equation}%
Meanwhile, the energy difference between two wave packets
\begin{eqnarray}
\Delta E &=&\frac{\Omega }{2}\left\langle \phi \left( n_{0},k_{0}\right)
\right\vert H_{\mathrm{0}}\left\vert \phi \left( n_{0},k_{0}\right)
\right\rangle  \nonumber \\
&&-\frac{\Omega }{2}\left\langle \phi \left( n_{0},k_{0}+\pi \right)
\right\vert H_{\mathrm{0}}\left\vert \phi (n_{0},k_{0}+\pi )\right\rangle
\nonumber \\
&=&2\Omega g\sqrt{\bar{n}}\cos k_{0}
\end{eqnarray}%
suppresses this process. It indicates that only under the condition $%
k_{0}\approx \pm \pi /2,$ such a transition has a higher probability to
occur due to the fact $\Delta E\approx 0$. We note that the condition $%
k_{0}\approx \pm \pi /2$\ is satisfied when $e^{-\mathrm{i}H_{\mathrm{0}%
}t}\left\vert \phi \left( n_{0},k_{0}\right) \right\rangle $ and $e^{-%
\mathrm{i}H_{\mathrm{0}}t}\left\vert \phi (n_{0},k_{0}+\pi )\right\rangle $\
overlap with each other in real space. The underlying mechanism of this
phenomenon is that $H_{\mathrm{1}}$ is a local interaction operator.\ To
demonstrate this point, numerical simulation is performed by exact
diagonalization for the truncated matrix of the Hamiltonian $H_{\mathrm{eq}}$
rather than $H_{\mathrm{eff}}$. The probabilities $P_{a}\left( t\right) $
and $P_{b}\left( t\right) $ are computed by $\left\vert \left\langle \psi
\left( 0\right) \right. e^{\mathrm{i}H_{\mathrm{0}}t}e^{-\mathrm{i}H_{%
\mathrm{eq}}t}\left\vert \psi \left( 0\right) \right\rangle \right\vert ^{2}$%
\ and $\left\vert \left\langle \psi \left( 0\right) \right. e^{\mathrm{i}H_{%
\mathrm{0}}\left( t+T_{B}/2\right) }e^{-\mathrm{i}H_{\mathrm{eq}%
}t}\left\vert \psi \left( 0\right) \right\rangle \right\vert ^{2}$\
respectively. In Fig. \ref{figure2} we plot the center positions and the
probability distributions of two evolved wave packets $e^{-\mathrm{i}H_{%
\mathrm{0}}t}\left\vert \psi \left( 0\right) \right\rangle $ and $e^{-%
\mathrm{i}H_{\mathrm{0}}\left( t+T_{B}/2\right) }\left\vert \psi \left(
0\right) \right\rangle $, obtained by analytical expression in Eqs. (\ref%
{n_ab}) and (\ref{p_ab}), and the numerical simulation.

It shows that the probability distributions are step-like as functions of
time and the probability exchange happens when two wave packets meet
together. When they meet each other for the first time, the splitting of a
Gaussian wave packet can be observed. It also indicates that our analytical
results, Eqs. (\ref{n_ab}) and (\ref{p_ab}), is the time-averaged
approximation for the probability transition.

A Gaussian wave packet with the center position $\bar{n}$ and the half width
$\Delta =2\sqrt{2\ln 2}/\alpha =23.5$, considering the evolution which has
been discussed above, can be regarded as being local within a spatial
extent, $D=L+6/\alpha $, where $L=-4g\sqrt{\bar{n}}/\omega $. When $\bar{n}%
\sim 10^{4}$, $L\sim 28$, $D<100$, now we let $D=100$; in this case, if we
want to satisfy $\left\vert \left( \sqrt{i}-\sqrt{\bar{n}}\right) /\sqrt{%
\bar{n}}\right\vert <5\%$ for $i\in D$, $\bar{n}\sim 10^{3}$ is mandatory at
least.

In the next section, the validity of the prediction will be investigated in
the Rabi system by numerical simulation. We will quantitatively evaluate the
extent of approximation of the above analysis.

\section{Controllable dynamics of cavity-atom system}

\label{sec_Controllable dynamics}Now we apply the obtained results to a
concrete case and then demonstrate the dynamic property of the system. We
investigate the time evolution of the wave packet in the single-mode
atom-cavity model. As mentioned above, the dynamics of local state in large
excitation number region is equivalent to that of the Hamiltonian in Eq. (%
\ref{H_eff_1}). We will show that the modulation of the Bloch oscillations
can be controlled by the coupling constant and the energy level of the atom,
which can be adjusted via external field.

As an application of the obtained result, let us take a simple case as an
example. Without loss of generality, we assume that the initial state is in
the form
\begin{equation}
\left\vert \psi \left( 0\right) \right\rangle =\frac{\left\vert
g\right\rangle +\left\vert e\right\rangle }{\sqrt{2R}}\sum_{n}e^{-\frac{%
^{\alpha ^{2}}}{2}\left( n-\bar{n}\right) ^{2}}\left\vert n\right\rangle .
\label{GWP1}
\end{equation}%
Transforming the basis $\left\{ \left\vert \mathrm{g},n\right\rangle
,\left\vert \mathrm{e},n\right\rangle \right\} $ to $\left\{ \left\vert
i\right\rangle _{\mathrm{e}}\right\} $ and $\left\{ \left\vert
i\right\rangle _{\mathrm{o}}\right\} $, we have%
\begin{equation}
\left\vert \psi \left( 0\right) \right\rangle =\frac{1}{\sqrt{2R}}%
\sum_{\lambda =\mathrm{e,o}}\sum_{i=0}^{\infty }e^{-\frac{^{\alpha ^{2}}}{2}%
\left( i-\bar{n}\right) ^{2}}\left\vert i\right\rangle _{\lambda },
\label{GWP2}
\end{equation}%
which corresponds to the superposition of two independent stationary wave
packets in two chains $H_{\mathrm{eff}}^{\lambda }$ with $\lambda =\mathrm{%
e,\ o}$, respectively. According to the above analysis, this state evolves to%
\begin{eqnarray}
\left\vert \psi \left( t\right) \right\rangle &=&\frac{1}{\sqrt{2R}}%
\sum_{n}^{\infty }e^{-\mathrm{i}n\omega t}\left\{ \left[ \vartheta _{t}\cos
\left( \gamma t\right) +\mathrm{i}\vartheta _{t+T_{B}/2}\sin \left( \gamma
t\right) \right] \left\vert g,n\right\rangle \right.  \nonumber \\
&&+\left. \left[ \vartheta _{t}\cos \left( \gamma t\right) -\mathrm{i}%
\vartheta _{t+T_{B}/2}\sin \left( \gamma t\right) \right] \left\vert
e,n\right\rangle \right\} ,
\end{eqnarray}%
where%
\begin{eqnarray}
\vartheta _{t} &=&\exp \left[ -\mathrm{i}\mu -\frac{^{\alpha ^{2}}}{2}\left(
n-n_{t}\right) ^{2}\right] ,  \label{theta_t} \\
\mu &=&-\frac{L}{2}\sin \left( \omega t\right) ,  \label{mu} \\
n_{t} &=&\bar{n}-L\sin ^{2}\left( \frac{\omega t}{2}\right) .  \label{n_t}
\end{eqnarray}%
We focus on the evaluation of the photon number distribution $P\left(
n,t\right) ,$\ characterizing the dynamics of the state. A straightforward
derivation shows that%
\begin{eqnarray}
P\left( n,t\right) &=&\left\vert \left\langle g,n\right. \left\vert \psi
\left( t\right) \right\rangle \right\vert ^{2}+\left\vert \left\langle
e,n\right. \left\vert \psi \left( t\right) \right\rangle \right\vert ^{2}
\nonumber \\
&=&\frac{1}{R}\left[ \left\vert \vartheta _{t}\right\vert ^{2}\cos
^{2}\left( \gamma t\right) +\left\vert \vartheta _{t+T_{B}/2}\right\vert
^{2}\sin ^{2}\left( \gamma t\right) \right]  \label{P_n}
\end{eqnarray}%
where $\left\vert \vartheta _{t}\right\vert ^{2}$ has the explicit form%
\begin{equation}
\left\vert \vartheta _{t}\right\vert ^{2}=\exp \left[ -\alpha ^{2}\left(
n-n_{t}\right) ^{2}\right] ,
\end{equation}%
which represents a Gaussian distribution with time-dependent center $n_{t}$.
It is shown that the photon number distribution $P\left( n,t\right) $\
exhibits amplitude modulated BOs.

We note that $P\left( n,t\right) $\ is directly determined by the factor $%
\gamma $ in Eq. (\ref{gamma}), which depends on the ratio $g\sqrt{\bar{n}}%
/\omega $. The numerical method is employed to simulate the time evolution
process. It is performed by exact diagonalization of the Hamiltonian with
truncated approximation. We are interested in the suppression and process of
the probability transition. In the following, we will investigate the cases
with two types of parameters: i) adjusting the ratio $g\sqrt{\bar{n}}/\omega
$\ for the zero-transition point, and ii) in-phase resonant oscillating $%
\Omega \left( t\right) $.

\subsection{Constant $\Omega $}

\begin{figure}[tbp]
\begin{center}
\includegraphics[bb=0 195 570 630, width=0.45\textwidth, clip]{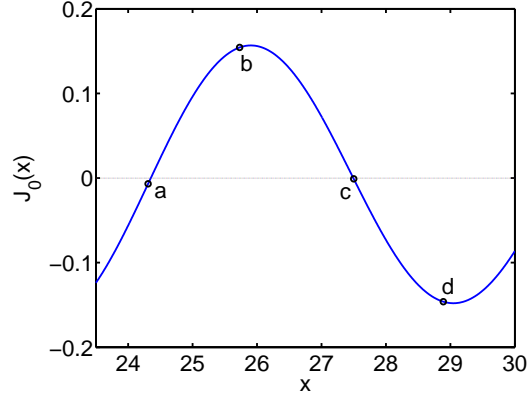}
\end{center}
\caption{(Color online) Bessel function $\mathrm{J}_{0}\left( x\right) $.
Two points a and c with $x=24.31$ and $27.50$, are in the vicinity of the
zeros. Points b and d with $x=25.73$ and $28.89$, are the stationary points.}
\label{figure3}
\end{figure}

\begin{figure}[tbp]
\begin{center}
\includegraphics[bb=20 170 580 520, width=0.45\textwidth, clip]{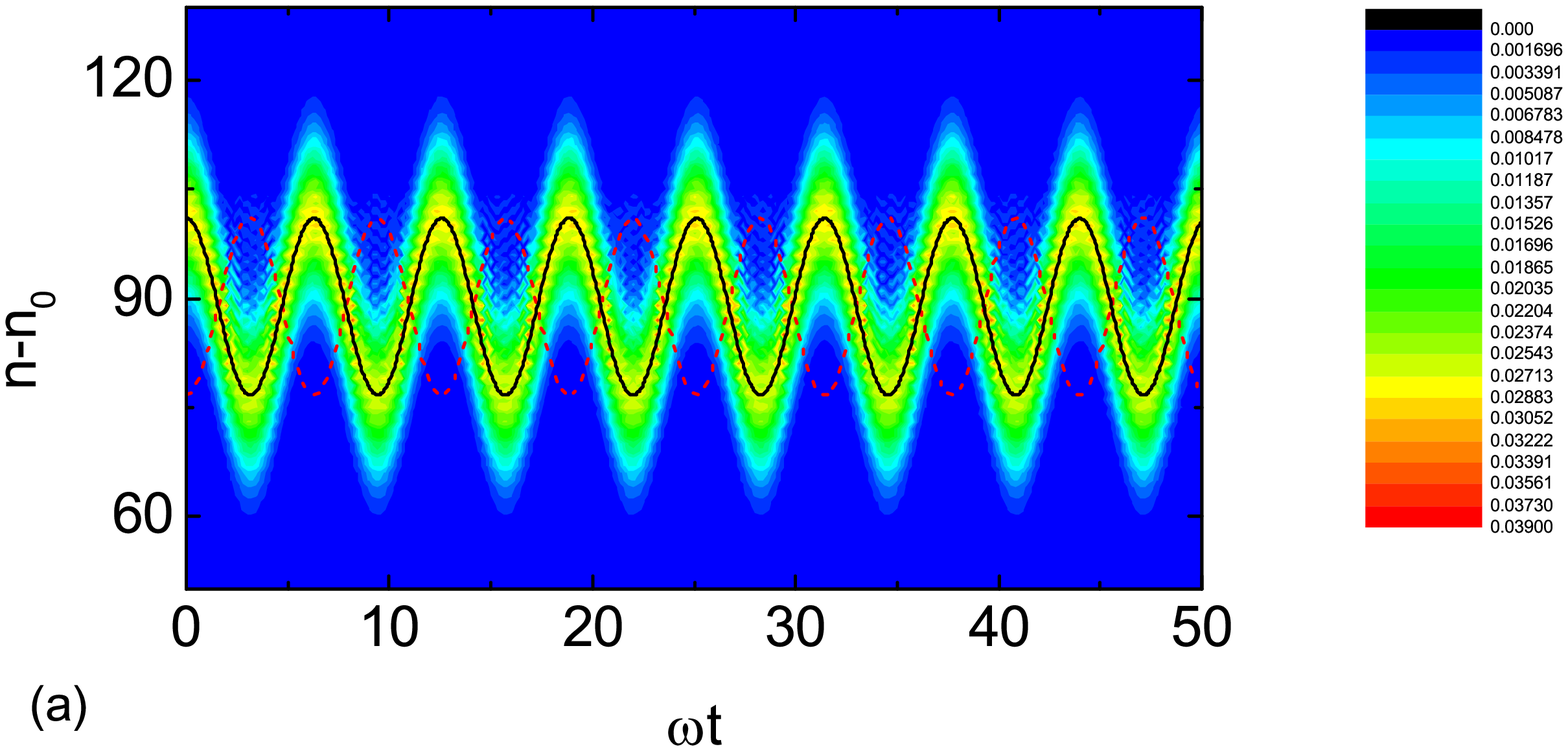} %
\includegraphics[bb=20 170 580 520, width=0.45\textwidth, clip]{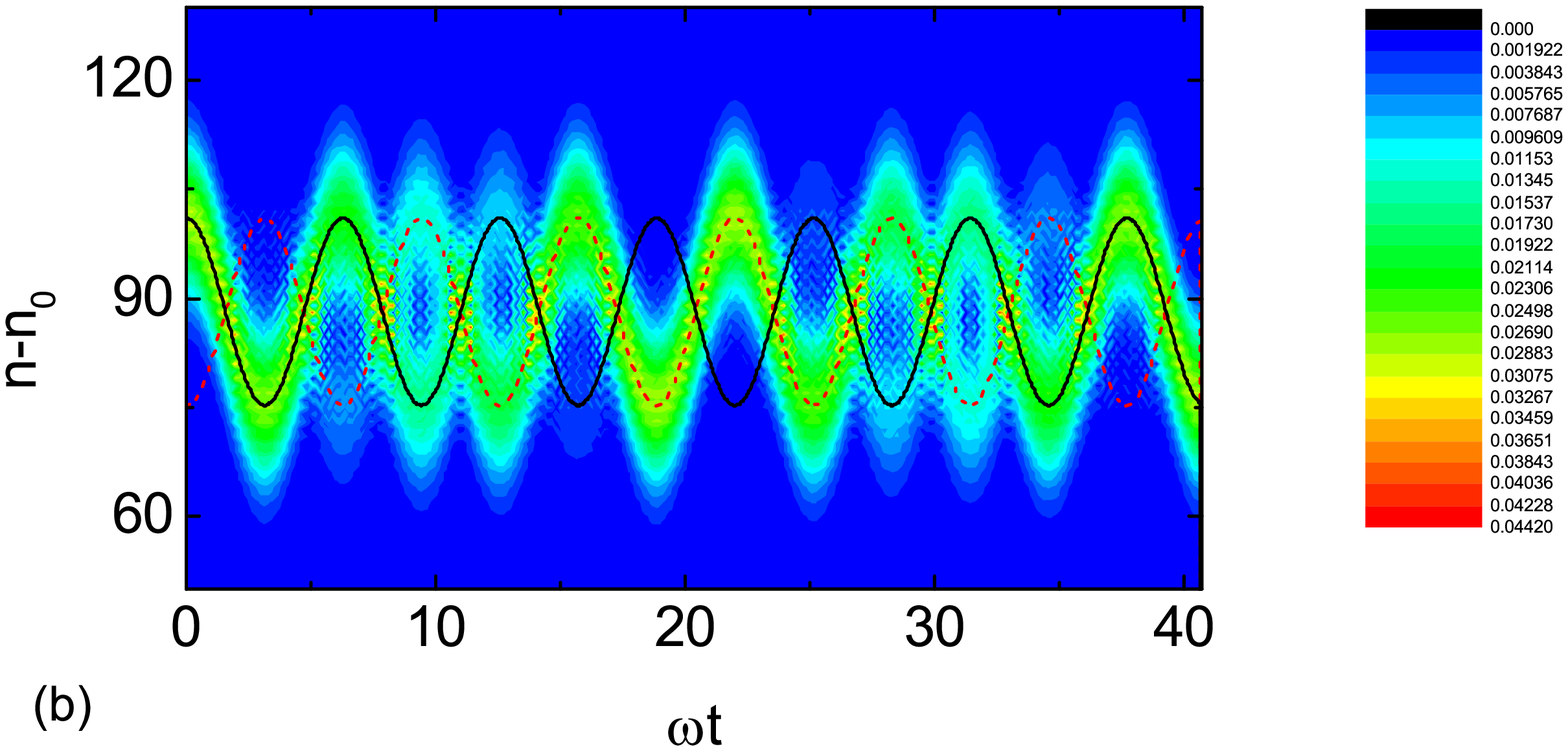} %
\includegraphics[bb=20 170 580 520, width=0.45\textwidth, clip]{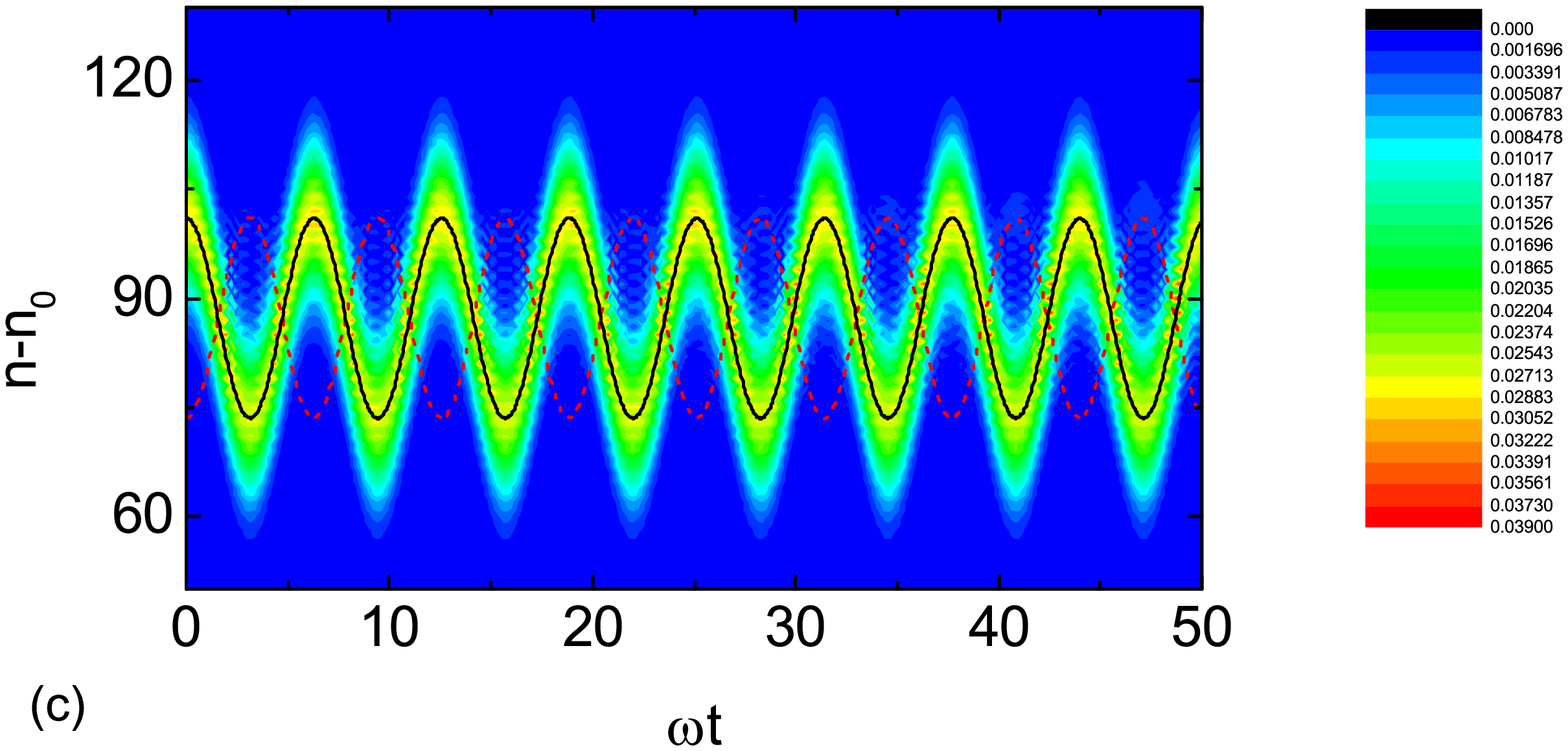} %
\includegraphics[bb=20 170 580 520, width=0.45\textwidth, clip]{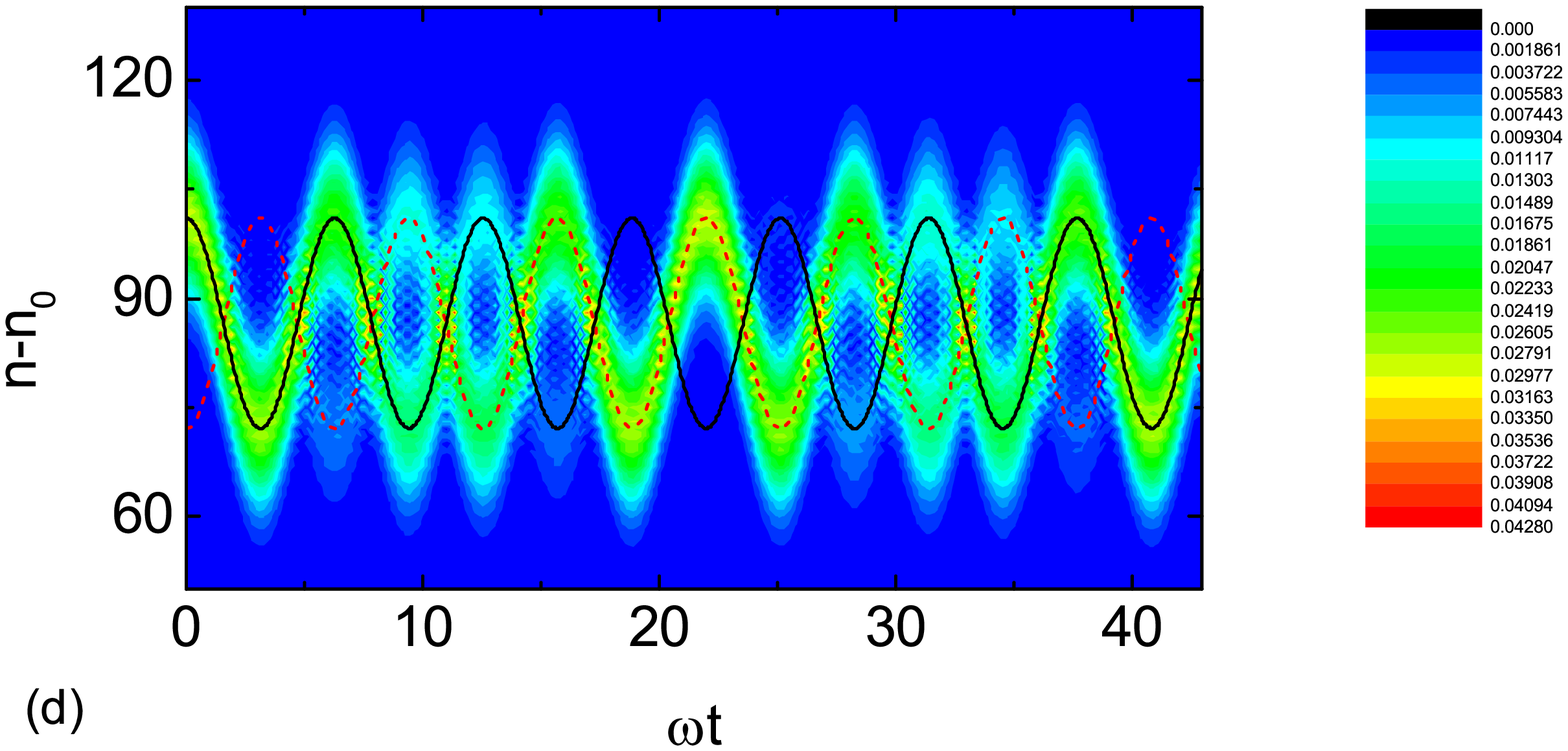}
\end{center}
\caption{(Color online) Time evolution of the initial state in Eq. (\protect
\ref{GWP1}) with the same $\bar{n}$\ and $\protect\alpha $\ as that in Fig. (%
\protect\ref{figure2}). It is driven by the original Rabi Hamiltonians with $%
\Omega =\protect\omega $ and parameters (see Fig. (\protect\ref{figure2}))
(a) $L =24.31$, (b) $25.73$, (c) $27.50 $, and (d) $28.89$. The photon
number distribution $P\left( n,t\right) $ is obtained by exact
diagonalization. The solid and dashed lines are the center positions of two
wavepackets from Eq. (\protect\ref{n_t}). It shows that the probability
transition is frozen in the cases of (a) and (c).}
\label{figure4}
\end{figure}

According to the approximate analysis, when we take $g\sqrt{\bar{n}}/\omega $
satisfying $\mathrm{J}_{0}\left( L\right) =0$, the transition between two
BOs can be frozen. However, in practice a deviation may occur with respect
to the approximate solution. We compute the time evolution and search the
zero-transition point for the initial state in the form of Eq. (\ref{GWP1}).
We consider four typical cases with $L=24.31$, $25.73$, $27.50$, and $28.89$%
. From the plot of the function $\mathrm{J}_{0}\left( x\right) $\ in Fig. (%
\ref{figure3}), we can see that two of the four points\ locate at\ the
vicinity of the zeros of the Bessel function,\ while the others locate at
the stationary points. In Fig. (\ref{figure4}), the photon number
distributions $P\left( n,t\right) $\ for these parameters are plotted. The
corresponding analytical results\textbf{\ }$n_{t}$\ \textbf{(}and\textbf{\ }$%
n_{t+T_{B}/2}$\textbf{) }in Eq. (\ref{n_t}) are also added on the plots of $%
P\left( n,t\right) $\ as a comparison and guide.

\subsection{In-phase resonant oscillating $\Omega \left( t\right) $\ }

We consider the resonant $\Omega \left( t\right) $, which have the half
period as the Bloch oscillations. To clarify the action of $\Omega \left(
t\right) $, we take a rectangular wave as%
\begin{equation}
\Omega \left( t\right) =\left\{
\begin{array}{cc}
\omega , &
\begin{array}{c}
\frac{1}{2}\left( nT-T_{1}\right) <\left\vert t-\varphi _{0}\right\vert \leq
\frac{1}{2}\left( nT+T_{1}\right) , \\
\left( n=1,2,3...,\right)%
\end{array}
\\
0 & \mathrm{otherwise}%
\end{array}%
\right.  \label{RW}
\end{equation}%
with $T=T_{B}$, $T_{1}=0.1T$. Here $\varphi _{0}$\ control the phase between
$\Omega \left( t\right) $\ and the Bloch oscillations. We consider two
typical cases with $\varphi _{0}=0$ and $T/4$, which are in- phase and
out-of-phase with the Bloch oscillations respectively.

\begin{figure}[tbp]
\begin{center}
\includegraphics[bb=50 30 600 630, width=0.45\textwidth,
clip]{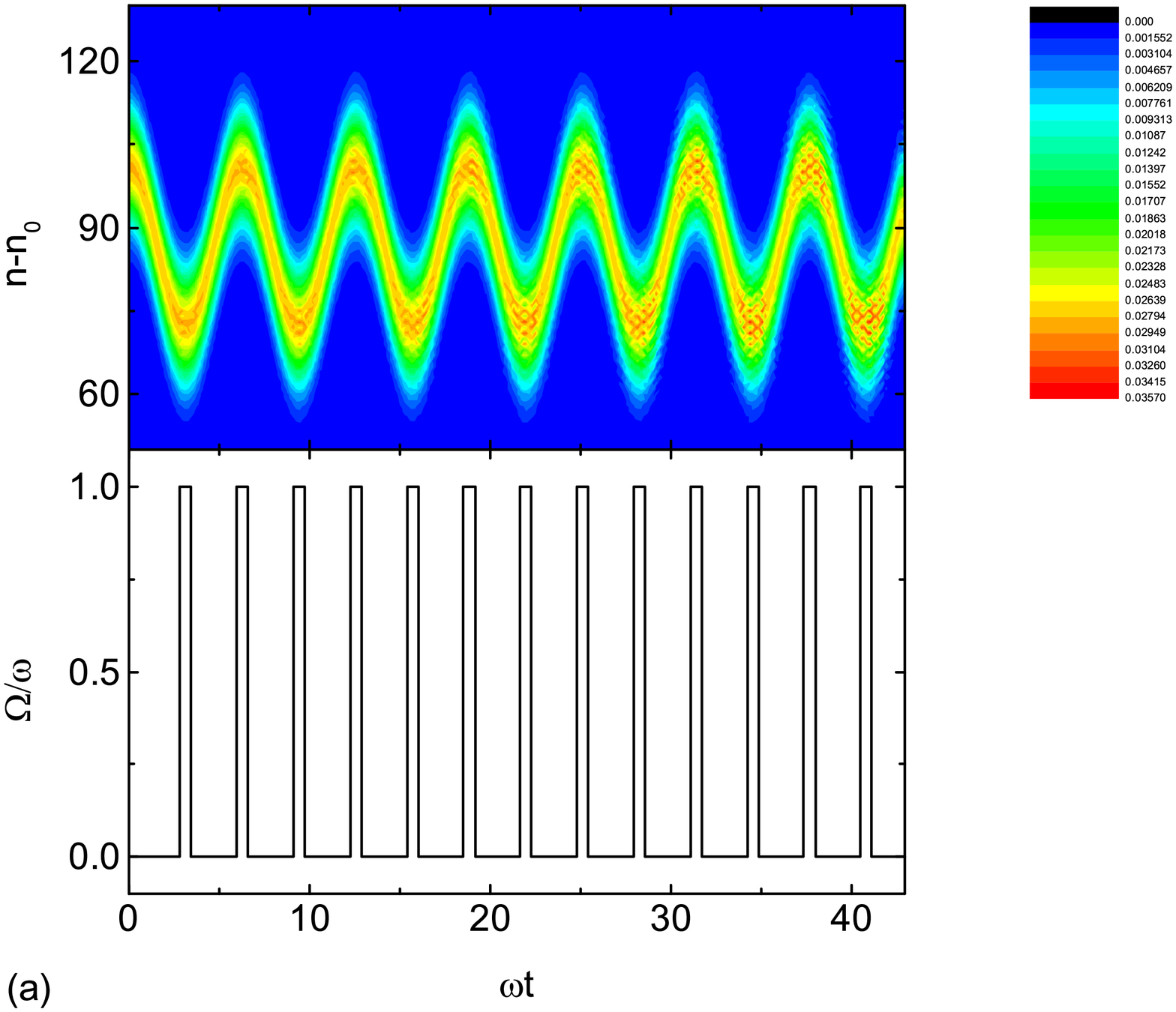}
\includegraphics[bb=50 30 600 630,
width=0.45\textwidth, clip]{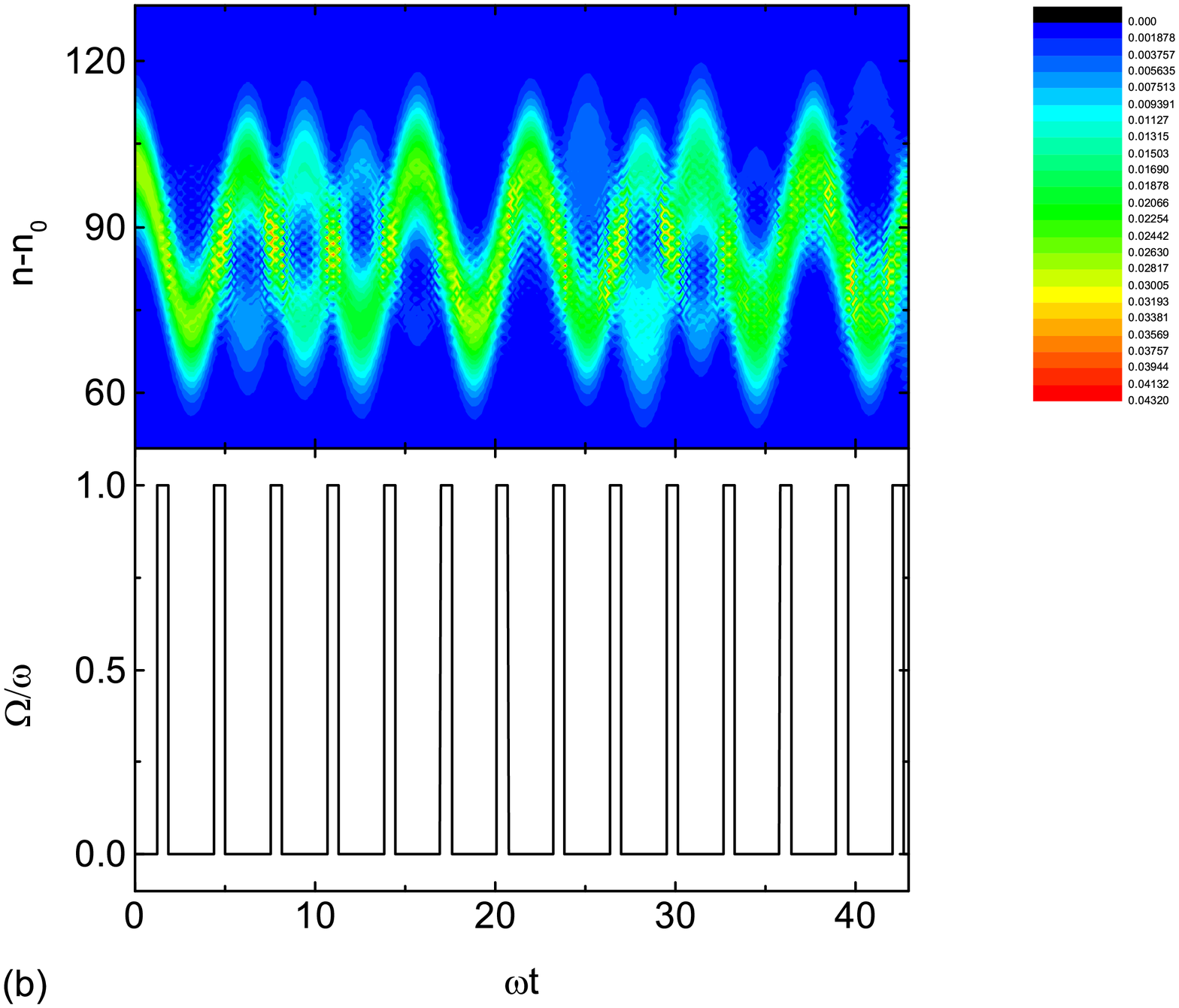}
\includegraphics[bb=50 30 600
630, width=0.45\textwidth, clip]{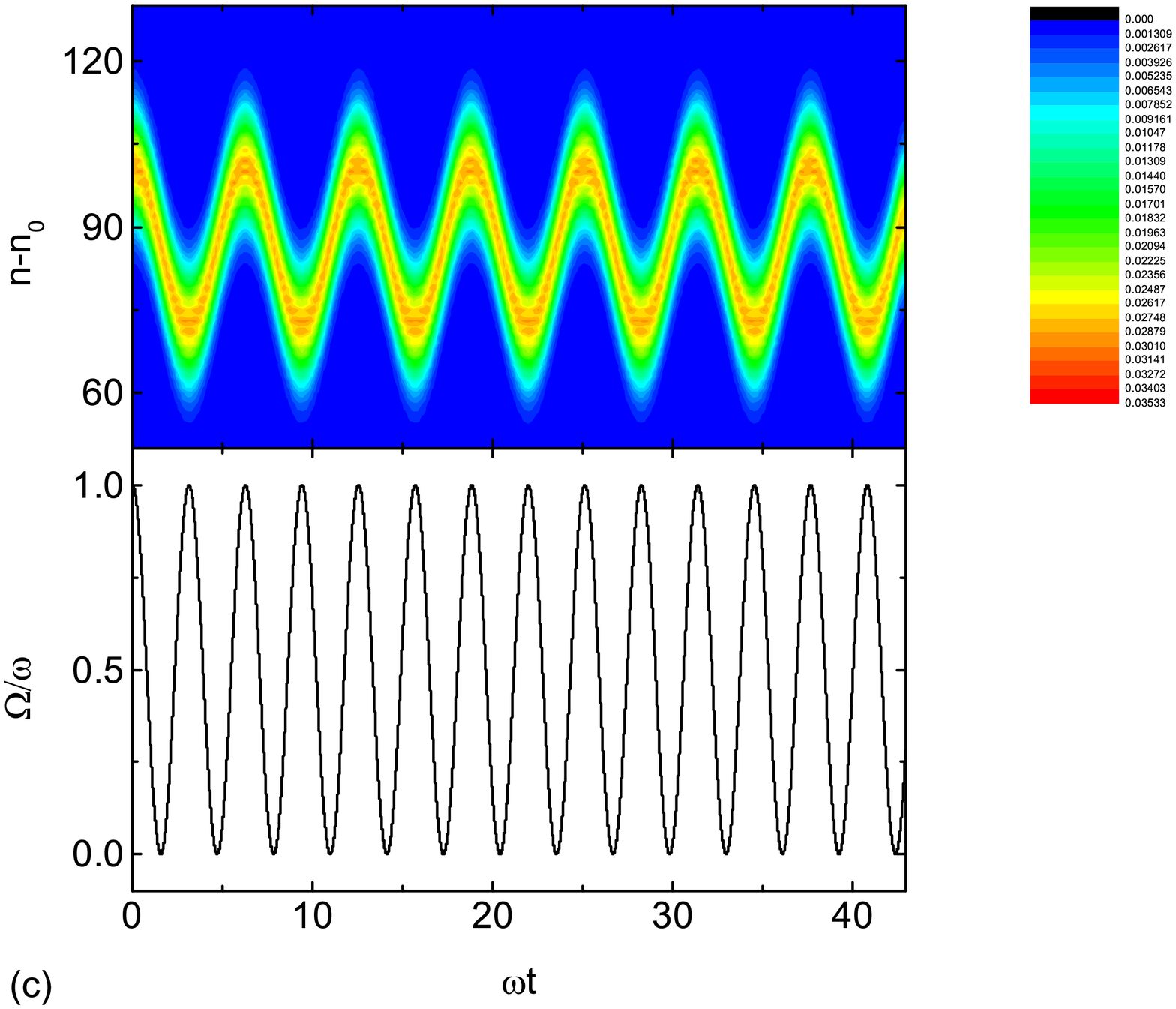}
\includegraphics[bb=50 30
600 630, width=0.45\textwidth, clip]{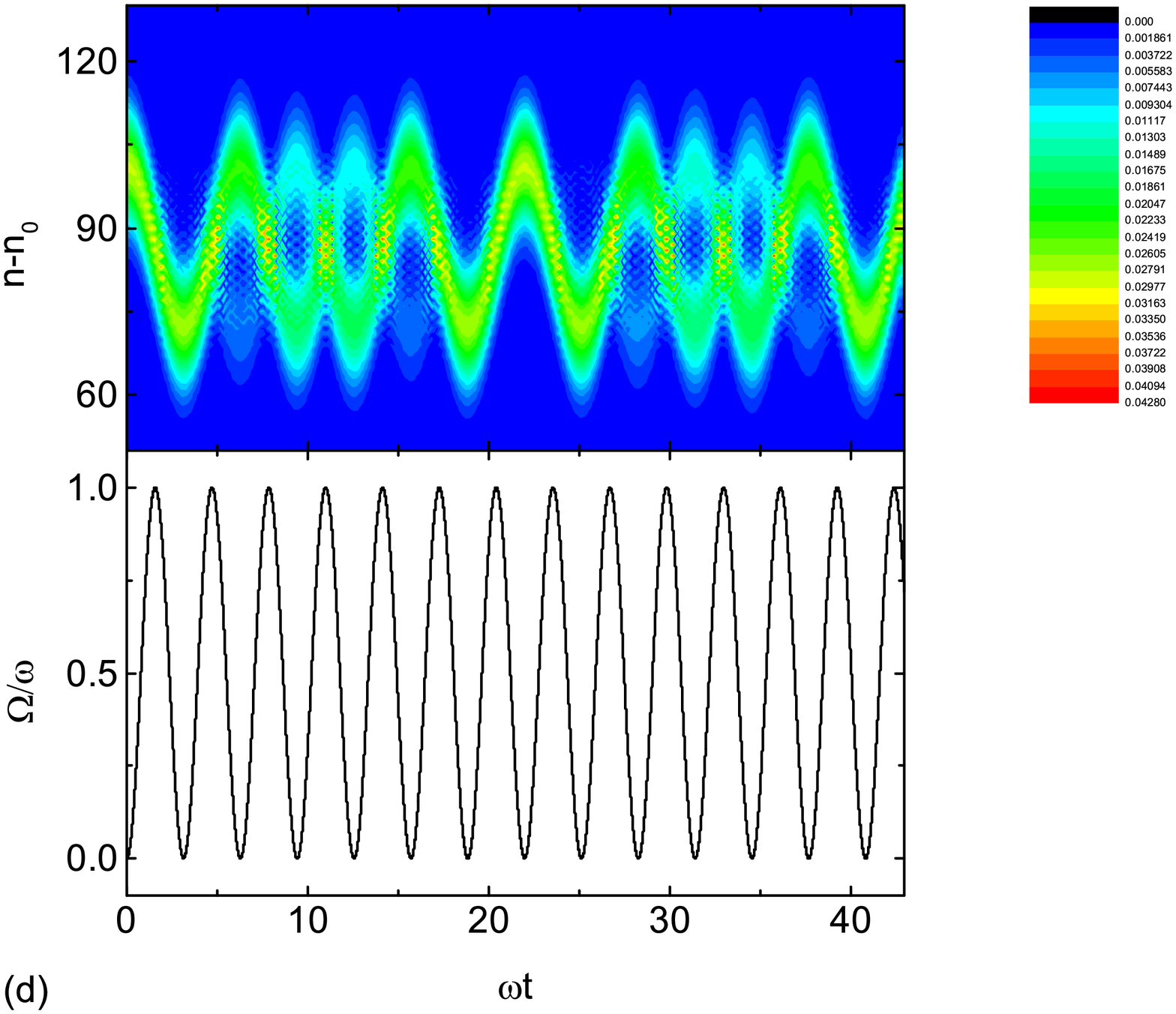}
\end{center}
\caption{(Color online) Time evolution of the initial state in Eq. (\protect
\ref{GWP1}) with the same $\bar{n}$\ and $\protect\alpha $\ as that in Fig. (%
\protect\ref{figure2}) $4g\protect\sqrt{\bar{n}}/\protect\omega =28.89$. The
$\Omega \left( t\right) $\ is taken in the form of Eq. (\protect\ref{RW})
with (a) $\protect\varphi _{0}=0$ and (b) $T/4$, and in the form of Eq. (%
\protect\ref{SW}) with (c) $\protect\varphi _{0}=0$ and (d) $T/4$. We can
see that for both square and sinusoidal waves of $\Omega \left( t\right) $,
the profile of the evolution in the case of in-phase (out-of-phase)\ is
similar to the case with constant $\Omega $\ (zero $\Omega $). This accords
with our prediction.}
\label{figure5}
\end{figure}

Similarly, we can also take a sinusoidal wave as
\begin{equation}
\Omega \left( t\right) =0.5\left\{ 1+\cos \left[ 2\omega \left( t+\varphi
_{0}\right) \right] \right\} \omega ,  \label{SW}
\end{equation}%
with $\varphi _{0}=0$ and $T/4$, respectively. Here we use a uniform mesh in
the time discretization for a time-dependent Hamiltonian, i.e.,%
\begin{equation}
\left\vert \psi \left( t\right) \right\rangle =\exp \left[
-i\int_{0}^{t}H\left( t^{\prime }\right) dt^{\prime }\right] \left\vert \psi
\left( 0\right) \right\rangle .
\end{equation}%
The profiles of the evolutions with different types of $\Omega \left(
t\right) $\ are plotted in Fig. \ref{figure5}. The numerical result accords
with our prediction: in the case of in-phase $\Omega \left( t\right) $,\ the
wave dynamics is similar to the case with constant $\Omega $, while in the
case of out-of-phase $\Omega \left( t\right) $, is similar to the case with
zero $\Omega $.

\section{Summary}

\label{sec_Summary}

In summary, we have studied the dynamics of the Rabi Hamiltonian,
identifying a parameter regime corresponding to amplitude modulated BOs,
another type of Bloch-Zener oscillations. It is the first time to present a
periodic phenomenon for this well-known\ model when the coupling constant
goes beyond the RWA. It also reveals a fact that a significant effect on the
field, Bloch-Zener oscillation of the photon probability distribution with a
distinct amplitude. It is remarkable that such a macroscopic phenomenon is
induced by a single atom. Moreover, the probability transition between the
two BOs can be controlled and suppressed by the ratio $g\sqrt{\bar{n}}%
/\omega $, as well as the in-phase resonant oscillating atomic frequency $%
\Omega \left( t\right) $, leading to multiple zero-transition\ points. The
numerical simulation of dynamics in the Rabi model confirms this prediction.
However, the experimental observation of this prediction requires a coherent
time scale of $g\sqrt{\bar{n}}$, which is still a challenge so far. At the
same time, with such a high number of photons in a quantum resonator,
dissipation may have a role, too.

\ack We acknowledge the support of the National Basic Research Program (973
Program) of China under Grant No. 2012CB921900 and CNSF (Grant No. 11374163).

\end{document}